\documentclass[prb,aps,twocolumn,showpacs,amsmath,amssymb,floatfix,citeautoscript]{revtex4}
\usepackage{graphicx}
\usepackage{dcolumn}
\usepackage{bm}
\usepackage{amsmath}
\usepackage{amssymb}
\usepackage[normal]{subfigure}

\usepackage{color}

\makeatletter
\begin{document}

\title{Tunneling Spectroscopy of a Spiral Luttinger Liquid in Contact with Superconductors}

\author{Dong E. Liu and Alex Levchenko}
\affiliation{Department of Physics and Astronomy, Michigan State
University, East Lansing, Michigan 48824, USA}

\date{September 25, 2013}

\begin{abstract}
One-dimensional wires with Rashba spin-orbit coupling, magnetic
field, and strong electron-electron interactions are described by a
spiral Luttinger liquid model. We develop a theory to investigate
the tunneling density of states into a spiral Luttinger liquid under
the proximity effect with superconductors. This approach provides a
way to disentangle the delicate interplay between superconducting
correlations and strong electron interactions. If the
wire-superconductor boundary is dominated by Andreev reflection, we
find that in the vicinity of the interface the zero-bias tunneling
anomaly reveals a power law enhancement with the unusual exponent.
Far away from the interface strong correlations inherent to the
Luttinger liquid prevail and restore conventional suppression of the
tunneling density of states at the Fermi level, which acquire,
however, a Friedel-like oscillatory envelope with the period
renormalized by the strength of the interaction.

\end{abstract}


\maketitle

\section{Introduction}

In a one-dimensional system strong electron-electron interactions
cause non-Fermi-liquid physics, which is described by the Luttinger
liquid theory~\cite{GiamarchiBook}. Nowadays quantum wires with
Rashba spin-orbit coupling in the external magnetic field attract a
great deal of attention due to their special charge and spin
transport, as well as spectral properties
\cite{Streta03,Sun-PRL07,BrauneckerPRL09,Schuricht-PRB12,
BrauneckerPRB12,Yacoby-ArXiv13,Loss-ArXiv13,Schmidt-ArXiv13}. The
presence of spin-orbit coupling leads to a relative shift of the
electronic dispersions for both spin species. Furthermore, a
magnetic field applied to the system lifts the spin degeneracy and
causes the opening of a Zeeman gap in the spectrum. If the chemical
potential lies inside the gap, the system is equivalent to a spinful
Luttinger liquid with a spiral magnetic field. This peculiar state
was abbreviated as a spiral Luttinger liquid (SLL). In the presence
of a bulk $s$-wave superconducting order, the wire becomes a
topological superconductor and hosts Majorana zero energy
modes~\cite{Lutchyn10,Oreg10}. Compelling evidence for the latter
has been recently reported in
experiments~\cite{kouwenhovenSCI12,Rokhinson12,deng12,das12}. It is
obviously of great interest to investigate the fate of
superconducting correlations when embedded into the environment of
the strongly interacting Luttinger liquid. Previous works considered
the phase diagram for this system in the presence of bulk
superconductivity~\cite{Gangadharaiah11,Sela11,Stoudenmire11}. Here
we develop a theory for the spatially and energy resolved tunneling
spectroscopy of a spiral Luttinger liquid which is brought into the
proximity to a superconductor (SC) at its boundary. This proposal
provides a way to disentangle the interplay between the complexity
of the superconducting effects and the nontrivial electron liquid
properties.

It is very well known from the context of mesoscopic conductors that
if the normal wire is placed between two superconductors, thus
forming a superconductor-normal-superconductor junction, its
spectral properties are strongly affected by the proximity
effect~\cite{Proximity-Review}. Indeed, the leakage of Cooper pairs
into the wire induces a nonvanishing superconducting pair amplitude
which opens a gap in the spectrum of the wire. For wires with a
length exceeding the superconducting coherence length, the gap is
small, of the order of Thouless energy $E_{Th}$, which evolves into
the complete superconducting gap $\Delta$ in the opposite limit of
short wires. If the normal wire is replaced by the Luttinger liquid
conductor, then even without superconducting perturbations the
density of states already has a striking feature. This is the famous
zero-bias anomaly -- the density of states vanishing as a power law
near the Fermi energy.~\cite{KF} One may naively expect that
proximitizing the Luttinger liquid with a superconductor would
further facilitate depletion of the states near the Fermi energy
towards opening a gap. Surprisingly, one discovers an entirely
different scenario, an enhancement of the anomaly -- the zero-bias
peak -- which physically can be rooted to the coherent
backscattering from the interface of the subgap excitations that
lead to the pileup of states near the zero
energy~\cite{Winkelholz96}. The latter has interesting consequences
for the Josephson effect in the Luttinger liquid constriction
between superconducting leads~\cite{Maslov96,Fazio,Smitha}. A
similar enhancement mechanism for tunneling has been also discussed
for a Luttinger liquid with impurity~\cite{Oreg-PRL96}. Here we
study this physics in the context of spiral Luttinger liquids.

The behavior of the tunneling density of states (TDOS) is sensitive
to the properties of the boundary between the Luttinger liquid and a
superconductor. Competition between normal and Andreev reflection
for this system has been discussed recently in the
literature~\cite{Fidkowski12}. We consider a perfect SC-SLL
interface that is dominated by an Andreev boundary
condition~\cite{Winkelholz96,Maslov96}. By performing a canonical
transformation to separate a gapless field from a gapped field, and
using a mode expansion, we obtain the low-energy asymptote for the
tunneling density of states analytically. We conclude that although
Zeeman splitting and a Rashba interaction destroy the TDOS
enhancement for the case when the chemical potential $\mu$ is
detuned from the Zeeman gap, an enhancement survives in the SLL
limit, namely, when $\mu$ lies within the gap. The power exponent of
this anomaly is different as compared to that in the conventional
case of a spinful Luttinger liquid without spin-orbit coupling. An
enhancement manifests only for distances close to the SC-SLL
interface in the wire, but disappears far away from the contact
where strong correlations inherent to SLL restore conventional power
law suppression of TDOS with additional oscillations. The latter
contribution is reminiscent of Friedel oscillations with the period
renormalized by interactions~\cite{Ussishkin}. We also compute the
tunneling density of states numerically by using a self-consistent
harmonic approximation and find the result to be consistent with the
analytical calculations.

\begin{figure}[t]
\centering
\includegraphics[width=2.7in,clip]{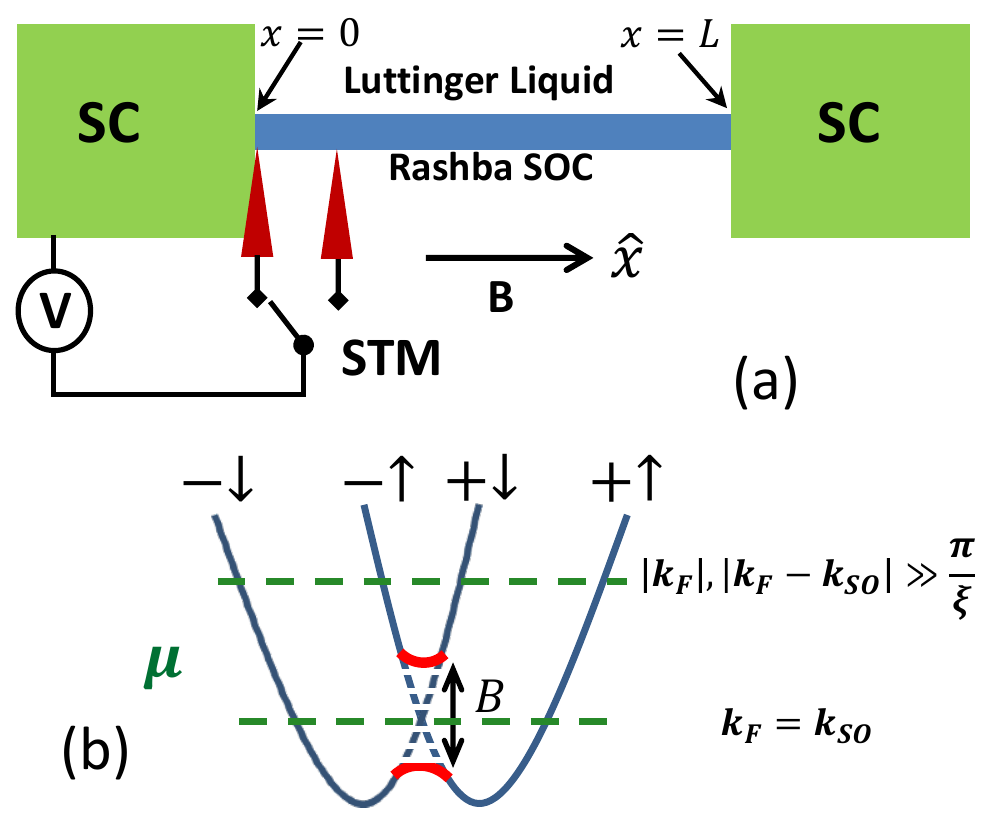}
\caption{(a) Schematic representation of a superconductor-spiral
Luttinger liquid wire system. (b) Band structure of the wire with
Rashba spin-orbit coupling (SOC) and a magnetic field. The green
dashed lines describe the position of the chemical potential $\mu$
for two different cases: (1) $|k_{F}|,\; |k_{F}-k_{\rm SO}| \gg
1/\xi$ for which $\mu$ is far above the Zeeman gap; (2)
$k_{F}=k_{\rm SO}$ for which $\mu$ lays within the Zeeman gap.}
\label{fig:setup}
\end{figure}

\section{Model and Hamiltonian}

We consider an interacting one-dimensional (1D) quantum wire with
Rashba spin-orbit coupling in the $z$ direction and a magnetic field
in parallel with the wire. Both sides of the wire are in perfect
contact with an $s$-wave superconductor. We are interested in the
density of states of the wire in the presence of electron-electron
interactions. A schematic setup of the system under consideration is
shown in Fig.~\ref{fig:setup}(a). The Hamiltonian of the central
part, i.e., the 1D Rashba wire with length $L$, can be written as
\begin{equation}
 H_{\rm W} = \int_{0}^{L} dx  \vec{\Psi}^{\dagger}(x) H(x) \vec{\Psi}(x)^{T}
           +\int_{0}^{L} dx dx' U \rho(x)\rho(x'),
\label{eq:H_wire}
\end{equation}
with
\begin{equation}
 H(x) = -\frac{\partial_x^2}{2 m} +\mu + B\hat{\sigma}_{x}
 -i\frac{k_{\rm SO}}{m}\hat{\sigma}_{z}\partial_x,
\end{equation}
where
$\vec{\Psi}^{\dagger}(x)=(\Psi_{\uparrow}^{\dagger}(x),\Psi_{\downarrow}^{\dagger}(x))$
being the electron annihilation operators for spin
$s=\uparrow,\downarrow$ at position $x$, and $k_{\rm SO}$ is the
Rashba spin-orbit momentum of the wire. The magnetic field is
applied along the $x$ direction, and is assumed to be uniform. The
second term in Eq.~(\ref{eq:H_wire}) represents the
electron-electron interaction with potential $U(x-x')$, and
$\rho(x)=\sum_{s}\Psi_{s}^{\dagger}(x)\Psi_{s}(x)$ is the electron
density (we choose $\hbar=1$ throughout the paper). It is convenient
to perform a spin-dependent gauge transformation,
$\Psi_{\uparrow\downarrow}(x)=\psi_{\uparrow\downarrow}(x)e^{\pm i
k_{\rm SO}x}$, followed by a standard bosonization in a linearized
spectrum:
\begin{eqnarray}
\hskip-.25cm \psi_{\pm,s}(x)\!\! &=&\!\! \frac{1}{\sqrt{2\pi\alpha}}
e^{i\sqrt{\frac{\pi}{2}}\Phi_{\pm,s}(x,t) } \nonumber\\
\hskip-.25cm &=&\!\! \frac{1}{\sqrt{2\pi\alpha}}
 e^{i\sqrt{\frac{\pi}{2}}
   \{\mp[\phi_{\rho}(x)+s\phi_{\sigma}(x)]+\theta_{\rho}(x)+s\theta_{\sigma}(x) \}}
\label{eq:bosonization}
\end{eqnarray}
with the full fermion operator $\psi_{s}(x) = e^{i k_F
x}\psi_{+,s}(x) + e^{-i k_F x} \psi_{-,s}(x)$, where
$\psi_{\pm,s}(x)$ represent the right and left moving fields,
respectively, and $\alpha$ is a conventional short distance cutoff.
Note that this transformation also shifts the chemical potential
$\mu\rightarrow \mu+k_{\rm SO}^2/2m$. After these steps, the system
is reduced to an equivalent spiral Luttinger liquid
model,~\cite{BrauneckerPRL09,BrauneckerPRB12} which is written in
terms of bosonic spin ($\sigma$) and charge ($\rho$) fields and
reads explicitly as
\begin{eqnarray}
H_{\rm W}= \sum_{\nu=\rho,\sigma}  \frac{v_{\nu}}{2} \int_{0}^{L}
dx\left[g_{\nu} (\partial_x\theta_{\nu})^2
            +g^{-1}_{\nu}(\partial_{x}\phi_{\nu})^2\right]  \nonumber\\
        +\frac{B}{2\alpha} \int_{0}^{L} dx
         \cos\left[\sqrt{2\pi}(\phi_{\rho}+\theta_{\sigma})-2(k_{F}-k_{\rm SO})x\right],\label{eq:H_wire-SLL}
\end{eqnarray}
where $g_{\rho,\sigma}$ are the interaction parameters and
$v_{\rho,\sigma}$ are the renormalized Fermi velocities. The fast
oscillating terms on the scales $\sim2k_{\rm SO}x$ and
$\sim2(k_{F}+k_{\rm SO})x$ are neglected in
Eq.~\eqref{eq:H_wire-SLL}. We discuss two possible cases as shown in
Fig. \ref{fig:setup} (b): (1) $\mu$ is far above the Zeeman gap,
i.e., $|k_{F}|,\; |k_{F}-k_{\rm SO}| \gg 1/\xi$ with the correlation
length being the minimal scale between the wire length and thermal
length $\xi=\mathrm{min}\{L,v_F/T\}$; (2) the chemical potential
$\mu$ lies in the middle of the Zeeman gap, i.e., $k_{F}\approx
k_{\rm SO}$. For case (1), the term proportional to $B$ strongly
oscillates and thus is irrelevant in the renormalization group (RG)
sense. Therefore, the magnetic field can be neglected in the
low-energy limit from Eq.~\eqref{eq:H_wire-SLL}, and the model
becomes the SC-spinful LL wire system with, however, an extra term,
$\sim \int dx\cos[\sqrt{8\pi}\theta_{\sigma}]$, due to the pair
hopping processes~\cite{Sun-PRL07}. This term induces a spin gap and
totally destroys the anomalous enhancement of TDOS. This behavior is
in sharp contrast with the SC-spinful LL wire without spin-orbit
physics involved~\cite{Winkelholz96}. For case (2), the cosine is
only slowly oscillating and such a spatial modulation that is due to
$2(k_{F}-k_{\rm SO})x$ can be dropped out if $|k_{F}-k_{\rm SO}| <
1/\xi$. In that limit, the magnetic field $B$ is relevant and will
grow as energy decreases if $g_{\rho}+1/g_{\sigma}<2$
\cite{BrauneckerPRL09,BrauneckerPRB12}. Note that the pair hopping
processes are strongly suppressed due to the Zeeman gap. We will
mostly focus on case (2) in this paper.

\section{Mode expansion for Andreev boundary condition}

We assume that the Rashba wire-superconductor interfaces are very
clean such that the Andreev reflection is the dominant process at
both boundaries. To treat the interfaces in the deep subgap limit,
$\varepsilon\ll\Delta$ with SC gap $\Delta$, we apply the following
fermion fields matching the
condition~\cite{Fidkowski12,Winkelholz96,Maslov96}
$\psi_{+,s}(x=0,L)=\mp ie^{i
\chi_{1,2}}\psi_{-,-s}^{\dagger}(x=0,L)$, where $s$ stands for
spin-up and spin-down channels, respectively, and $\chi_{1,2}$ are
the phases of the SC order for the left (near $x=0$) and right (near
$x=L$) superconductors. It is important to emphasize that the
spin-dependent gauge transformation used above to transform the
Hamiltonian leaves invariant both the Cooper pairing term in the
$s$-wave SC and the Andreev boundary condition. To proceed, we adopt
the canonical mode expansion~\cite{Winkelholz96,Maslov96}
\begin{eqnarray}
 \theta_{\rho}(x)\!\! &=&\!\! \sqrt{\frac{\pi}{2}}(J+\chi)\frac{x}{L}+i\sqrt{\frac{1}{g_\rho}}
                    \sum_{q>0}\gamma_{q} \sin(qx)(b_{\rho q}^{\dagger}-b_{\rho q}),\nonumber\\
 \theta_{\sigma}(x)\!\! &=&\!\! \frac{\theta_{\sigma}^{0}}{\sqrt{\pi}}+\sqrt{\frac{1}{g_\sigma}}
                    \sum_{q>0}\gamma_{q} \cos(qx)(b_{\sigma q}^{\dagger}+b_{\sigma q}),\nonumber\\
\phi_{\rho}(x)\!\! &=&\!\!
\frac{\phi_{\rho}^{0}}{\sqrt{\pi}}+\sqrt{g_{\rho}}
                    \sum_{q>0}\gamma_{q} \cos(qx)(b_{\rho q}^{\dagger}+b_{\rho q}),\nonumber\\
 \phi_{\sigma}(x)\!\! &=&\!\! \sqrt{\frac{\pi}{2}}M\frac{x}{L}+i\sqrt{g_{\sigma}}
                    \sum_{q>0}\gamma_{q} \sin(qx)(b_{\sigma q}^{\dagger}-b_{\sigma q}),
\label{eq:ModeExpansion}
\end{eqnarray}
where $\chi=\chi_1-\chi_2$ is the global phase difference between
two SC islands, $b_{\rho q}$ and $b_{\sigma q}$ are bosonic
operators, $\gamma_{q}=e^{-q\alpha/2\pi}/\sqrt{qL}$ is the
convergence factor, and $q=\pi n/L$ ($n=1,2,...$). The zero mode
operators, satisfying the commutations $[\theta_{\sigma}^{0},\,M]=i$
and $[\phi_{\sigma}^{0},\,J]=i$, describe the topological
excitations~\cite{Haldane}. Note that those are not the eigenmodes
for our system, and just serve as the starting point for the
diagonalization later.

\section{Low-energy TDOS of the wire}

The magnetic field will flow to strong coupling for
$g_{\rho}+1/g_{\sigma}<2$ at low energy. To separate the
corresponding gapped field from a gapless part, one can apply the
following canonical transformation:~\cite{BrauneckerPRL09}
\begin{eqnarray}
 \phi_{\rho}=\frac{g_{\rho}}{\sqrt{g}}\,\phi_{+}+\sqrt{\frac{g_{\rho}}{g_{\sigma}g}}\,\phi_{-}\,,\;
   \theta_{\rho}=\frac{1}{\sqrt{g}}\,\theta_{+}+\frac{1}{\sqrt{g_{\rho}g_{\sigma}g}}\,\theta_{-}\nonumber\\
 \phi_{\sigma}=\frac{1}{\sqrt{g}}\,\theta_{+}-\sqrt{\frac{g_{\rho}}{g_{\sigma}g}}\,\theta_{-}\,,\;
   \theta_{\sigma}=\frac{1}{g_{\sigma}\sqrt{g}}\,\phi_{+}-\sqrt{\frac{g_{\rho}}{g_{\sigma}g}}\,\phi_{-}\,,
\label{eq:CT}
\end{eqnarray}
where $g=g_{\rho}+1/g_{\sigma}$. The Hamiltonian
\eqref{eq:H_wire-SLL} then becomes
\begin{eqnarray}
 H_{W}&=&\int_{0}^{L}dx \sum_{i=\pm}\frac{u_{i}}{2}
        \Big[ (\partial_x \theta_{i})^2 + (\partial_x \phi_{i})^2  \Big] \nonumber\\
      & &+ \frac{B}{2\alpha}\int_{0}^{L}dx \cos[\sqrt{2\pi g} \phi_{+}(x)],
\end{eqnarray}
where $u_{+}=(v_{\rho}g_{\rho}+v_{\sigma}/g_{\sigma})/g$ and
$u_{-}=(v_{\rho}/g_{\sigma}+v_{\sigma}g_{\rho})/g$. The off-diagonal
terms $\sim (\partial_x\phi_{+})(\partial_x\phi_{-})$ and $\sim
(\partial_x\theta_{+})(\partial_x\theta_{-})$ are neglected in a
mean-field treatment for large $B$
\cite{BrauneckerPRL09,Loss-ArXiv13}. This canonical transformation
along with Eq.~(\ref{eq:ModeExpansion}) results in:
\begin{eqnarray}
\theta_{+}(x)\!\! &=&\!\! \sqrt{\frac{\pi}{2}}N_{+}\frac{x}{L}
                + i \sum_{q>0} \gamma_{q} \sin(qx) (b_{+q}^{\dagger}-b_{+q}),\nonumber\\
\theta_{-}(x)\!\! &=&\!\! \sqrt{\frac{\pi}{2}}
\sqrt{\frac{g_{\rho}}{g_{\sigma}}} N_{-}\frac{x}{L}
                + i \sum_{q>0} \gamma_{q} \sin(qx) (b_{-q}^{\dagger}-b_{-q}),\nonumber\\
\phi_{+}(x)\!\! &=&\!\! \frac{\phi_{+}^{(0)}}{\sqrt{\pi}}
                + \sum_{q>0} \gamma_{q} \cos(qx) (b_{+q}^{\dagger}+b_{+q}),\nonumber\\
\phi_{-}(x)\!\! &=&\!\!
\sqrt{\frac{g_{\sigma}}{g_{\rho}}}\frac{\phi_{-}^{(0)}}{\sqrt{\pi}}
                + \sum_{q>0} \gamma_{q} \cos(qx) (b_{-q}^{\dagger}+b_{-q}),
\label{eq:ME_PM}
\end{eqnarray}
where $b_{\pm,\,q}=(\pm\sqrt{g_{\rho}}b_{\rho/\sigma,\,
q}+b_{\sigma/\rho,\, q}/\sqrt{g_{\sigma}})/\sqrt{g}$,
$\phi_{+}^{(0)}=(\phi_{\rho}^{0}+\phi_{\sigma}^{0})/\sqrt{g}$,
$\phi_{-}^{(0)}=(\phi_{\rho}^{0}/g_{\sigma}-g_{\rho}\phi_{\sigma}^{0})/\sqrt{g}$,
$N_{+}=(M/g_{\sigma}+g_{\rho}(J+\chi))/\sqrt{g}$, and
$N_{-}=(J+\chi-M)/\sqrt{g}$.

\begin{figure}[t]
\centering
\includegraphics[width=3.3in,clip]{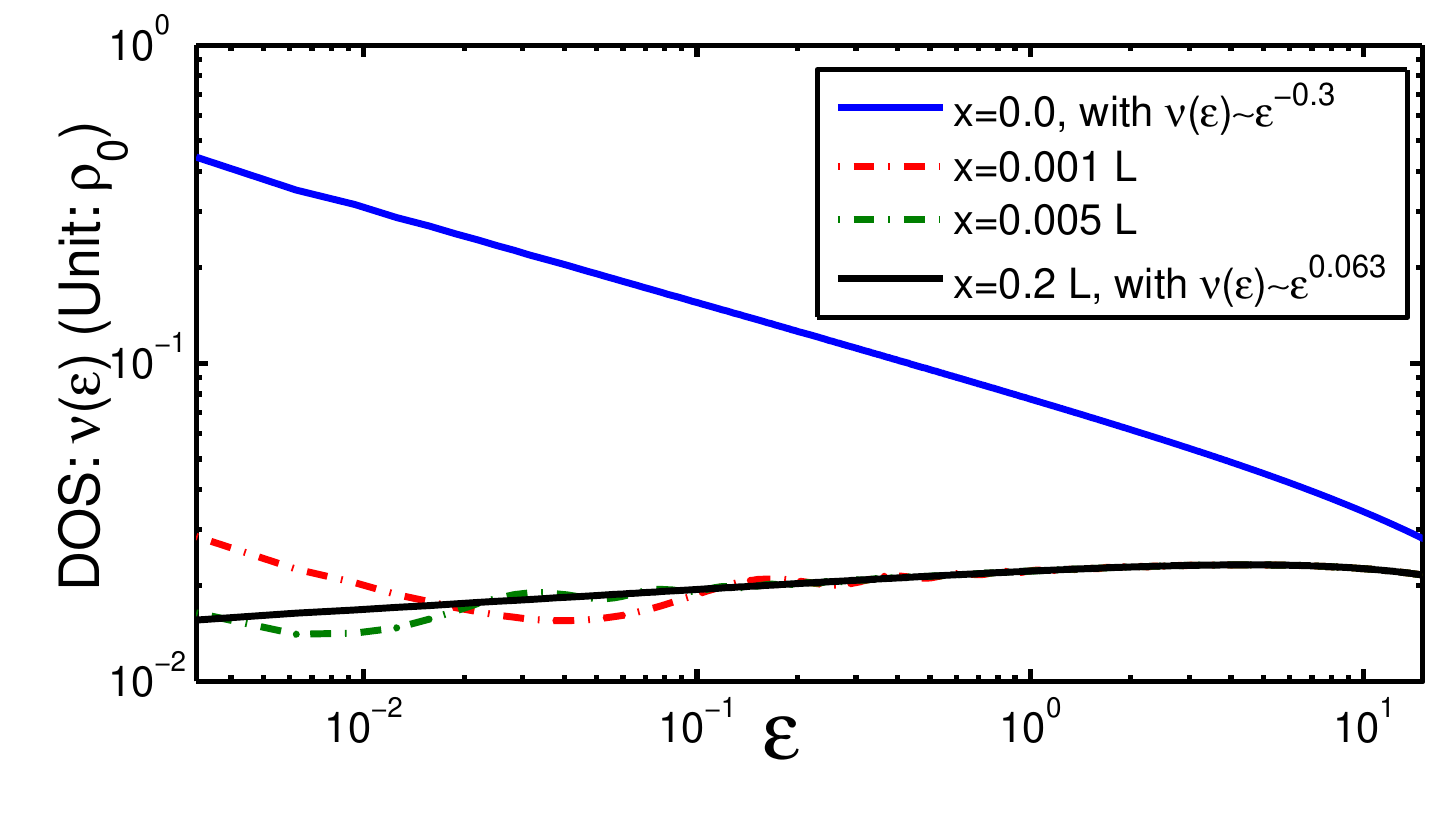}
\caption{The TDOS for a finite long wire using Eq. (\ref{eq:CORR1})
and the mode expansion Eq. (\ref{eq:ME_PM}). The parameters are:
$L=10000$, $\alpha=0.01$, $g_{\rho}=0.5$, $g_{\sigma}=0.85$,
$v_{\rho}=0.8$, $v_{\sigma}=0.47$, and
$\rho_0=1/\sqrt{2\pi\alpha}$.} \label{fig:AnalyticFig}
\end{figure}

The density of states in a wire measured at a distance $x$ from the
left interface is given by the Fourier transform of the retarded
Green's function
$\mathcal{G}^R(x,x',t)=-i\theta(t)\langle\{\Psi(x,t),\Psi^{\dagger}(x',0)
\}\rangle$,
\begin{equation}
 \nu(x,\varepsilon)=-\frac{1}{\pi}\Im\int_{-\infty}^{+\infty}
 dt e^{i\varepsilon t}\mathcal{G}^R(x,x,t). \label{eq:DOSF}
\end{equation}
Here, $\Psi(x,t)=\Psi_{\uparrow}(x,t)+\Psi_{\downarrow}(x,t)$ and
$\Psi_{s}(x,t)=e^{i(k_{F}+s k_{\rm SO})}\psi_{+,s}(x,t)+
e^{i(-k_{F}+s k_{\rm SO})}\psi_{-,s}(x,t)$, where $\psi_{\pm,s}$ is
obtained using Eq.~(\ref{eq:bosonization}) and the mode expansion
Eq.~\eqref{eq:ME_PM}. The correlation function includes the
following terms
\begin{eqnarray}
 &\langle \Psi(x,t) \Psi^{\dagger}(x,0) \rangle = \sum_{\alpha=\pm,s} \langle \psi_{\alpha,s}(x,t)
                \psi^{\dagger}_{\alpha,s}(x,0) \rangle \nonumber\\
 &\quad +\langle \psi_{-\uparrow}(x,t)\psi^{\dagger}_{+\downarrow}(x,0)  \rangle
    +\langle \psi_{+\downarrow}(x,t)\psi^{\dagger}_{-\uparrow}(x,0)
    \rangle.
\label{eq:FC}
\end{eqnarray}
Some other terms are zero due to the neutrality condition, i.e.,
$\langle e^{i A \phi_{-}^{(0)}} \cdots \rangle=0$ for $A\neq 0$.
Note that this is not true for $\langle e^{i A \phi_{+}^{(0)}}
\cdots \rangle$ due to the cosine potential. In the low-energy
limit, $\phi_{+}(x,t)$ is pinned to the local minima of the cosine
potential and behaves as a constant phase. Using the canonical
transformation shown in Eq.~(\ref{eq:CT}) and dropping out the
constant phase, the bosonized field $\Phi_{\pm,s}(x,t)$ introduced
in Eq.~(\ref{eq:bosonization}) becomes
\begin{eqnarray}
 \Phi_{+,\uparrow}\!\! &=&\!\! -2\sqrt{\frac{g_{\rho}}{g_{\sigma}g}}\, \phi_{-}
   +\left(\sqrt{\frac{g_{\rho}g_{\sigma}}{g}}+\frac{1}{\sqrt{g_{\rho}g_{\sigma}g}}\right)\theta_{-},\nonumber\\
 \Phi_{+,\downarrow}\!\! &=&\!\! \left(\frac{1}{\sqrt{g_{\rho}g_{\sigma}g}}-
 \sqrt{\frac{g_{\rho}g_{\sigma}}{g}}\right)\theta_{-}(x,t)
              +\frac{2}{\sqrt{g}}\theta_{+}, \nonumber\\
 \Phi_{-,\uparrow}\!\! &=&\!\!\left(\frac{1}{\sqrt{g_{\rho}g_{\sigma}g}}-\sqrt{\frac{g_{\rho}g_{\sigma}}{g}}\right)\theta_{-}(x,t)
              +\frac{2}{\sqrt{g}}\theta_{+},\nonumber\\
 \Phi_{-,\downarrow}\!\! &=&\!\! 2\sqrt{\frac{g_{\rho}}{g_{\sigma}g}}\, \phi_{-}
              +\left(\sqrt{\frac{g_{\rho}g_{\sigma}}{g}}+\frac{1}{\sqrt{g_{\rho}g_{\sigma}g}}\right)\theta_{-}.
\end{eqnarray}
The dual field $\theta_{+}(x,t)$ is totally disordered and the
correlation $\langle e^{i\alpha\theta_{+}(x,t)}
e^{-i\alpha\theta_{+}(x,0)} \rangle$ decays exponentially to zero
(as a function of $t$), and therefore any term in the density of
states including such correlations does not show a power law
divergence, which can be safely neglected for our purpose. Then, the
correlation function can be simplified to
\begin{eqnarray}
\langle \Psi(x,t) \Psi^{\dagger}(x,0) \rangle =
\langle \psi_{+\uparrow}(x,t)\psi^{\dagger}_{+\uparrow}(x,0)  \rangle \nonumber\\
+\langle \psi_{-\downarrow}(x,t)\psi^{\dagger}_{-\downarrow}(x,0)
\rangle. \label{eq:CORR1}
\end{eqnarray}
By using now Eqs.~(\ref{eq:bosonization}), (\ref{eq:CT}), and
(\ref{eq:ME_PM}), the correlation functions for a finite wire and
for $\chi=0$ (condition of the absence of the supercurrent) yield
\begin{eqnarray}
 \hskip-1cm&\langle \Psi(x,t) \Psi^{\dagger}(x,0) \rangle = \frac{1}{\pi\alpha}
   \Bigg[\frac{1-e^{-\pi\alpha/L}}{1-e^{-\pi(i u_{-}t+\alpha)/L}}\Bigg]^{\eta+\beta}\nonumber\\
 \hskip-1cm&\quad\quad\times  \Bigg[\frac{\Big( 1-e^{-\pi(\alpha-2ix)/L}\Big)\Big( 1-e^{-\pi(\alpha+2ix)/L}\Big)}
        {\Big( 1-e^{-\pi(i(u_{-}t-2x)+\alpha)/L}\Big)
        \Big(1-e^{-\pi(i(u_{-}t+2x)+\alpha)/L}\Big)}\Bigg]^{\frac{\eta-\beta}{2}},
\end{eqnarray}
which, in the long wire limit, becomes
\begin{eqnarray}
 \langle \Psi(x,t) \Psi^{\dagger}(x,0) \rangle
  =\frac{1}{\pi\alpha}\left[\frac{\alpha}{i u_{-}t + \alpha}\right]^{\eta +
  \beta}\nonumber\\
  \times\left[
  \frac{\alpha^2+(2x)^2}{[i(u_{-}t-2x)+\alpha][i(u_{-}t+2x)+\alpha]}\right]^{\frac{\eta-\beta}{2}},
\label{eq:corr_Linf}
\end{eqnarray}
where $\eta=g_{\rho}/(1+g_\rho g_{\sigma})$ and $\beta=1/4\eta$.
Finally, performing a Fourier integral, we find that at low energy
$\varepsilon\ll \Delta$, the TDOS at the SC-SLL interface $x=0$
follows the unusual power law
\begin{equation}
\nu(0,\varepsilon)=\frac{2}{\pi\Gamma(2\eta)u_-}\left[\frac{\alpha\varepsilon}{u_-}\right]^{2\eta-1},
\label{eq:ADOS_interface}
\end{equation}
where $\Gamma$ is the Euler gamma function. Since we are in the
regime $g_{\rho}+1/g_{\sigma}<2$, such that the cosine term is
relevant, this power is always negative, $2\eta-1<0$, which induces
an anomalous density of states enhancement at the Fermi energy
(i.e., zero voltage bias peak). For $x\neq 0$, the TDOS is obtained
by integrating over $t$ along three branch cuts (with branching
points $i\alpha$ and $\pm2x/u_{-}+i\alpha$,) in the complex $t$
plane. In the limit $2x\varepsilon/u_{-}\gg 1$, the contributions of
those branch cuts can be calculated independently (see
Appendix~\ref{app:DOS} for further details). One then obtains the
density of states asymptote far from the interface $x\gg
u_-/\varepsilon$:
\begin{eqnarray}
\hskip-.75cm &&  \nu(x,\varepsilon) =
\frac{1}{\pi\Gamma(\eta+\beta)u_-}
  \left[\frac{\alpha\varepsilon}{u_-}\right]^{\eta+\beta-1}  \nonumber\\
\hskip-.75cm &&
+\frac{2^{2-\eta-\beta}\cos(2x\varepsilon/u_{-}+\delta)}{\pi\Gamma((\eta-\beta)/2)u_-}
   \left[\frac{\alpha\varepsilon}{u_-}\right]^{\frac{\eta-\beta}{2}-1}\!\!
   \left[\frac{\alpha}{x}\right]^{\frac{\eta}{2}+\frac{3\beta}{2}}\!\!, \label{eq:DOS_FiniteX}
\end{eqnarray}
where the phase shift is
$\delta=\mathrm{Arg}(i^{\frac{3\eta}{2}+\frac{\beta}{2}})$ and
$\eta+\beta-1>0$. Figure \ref{fig:AnalyticFig} represents TDOS for a
finite long wire computed numerically from Eqs.~(\ref{eq:ME_PM}) and
(\ref{eq:CORR1}). Here, we choose a finite frequency resolution in
the numerical Fourier transformation. For $x=0$ and a specific
choice of the interaction parameters indicated in the caption of
Fig.~\ref{fig:AnalyticFig}, the TDOS displays a clear power law
enhancement at zero energy: $\nu\propto \varepsilon^{-0.3}$. For
small $x$, one can see the oscillation. For large distances
($x=0.2L$ away from the interface), the factor
$x^{-\frac{\eta}{2}-\frac{3\beta}{2}}$ makes an oscillatory term
invisible in the plot, while the main contribution to TDOS shows a
power law decay $\nu\propto
\varepsilon^{\eta+\beta-1}\propto\varepsilon^{0.063}$.

\begin{figure}[t]
\centering
\includegraphics[width=3.1in,clip]{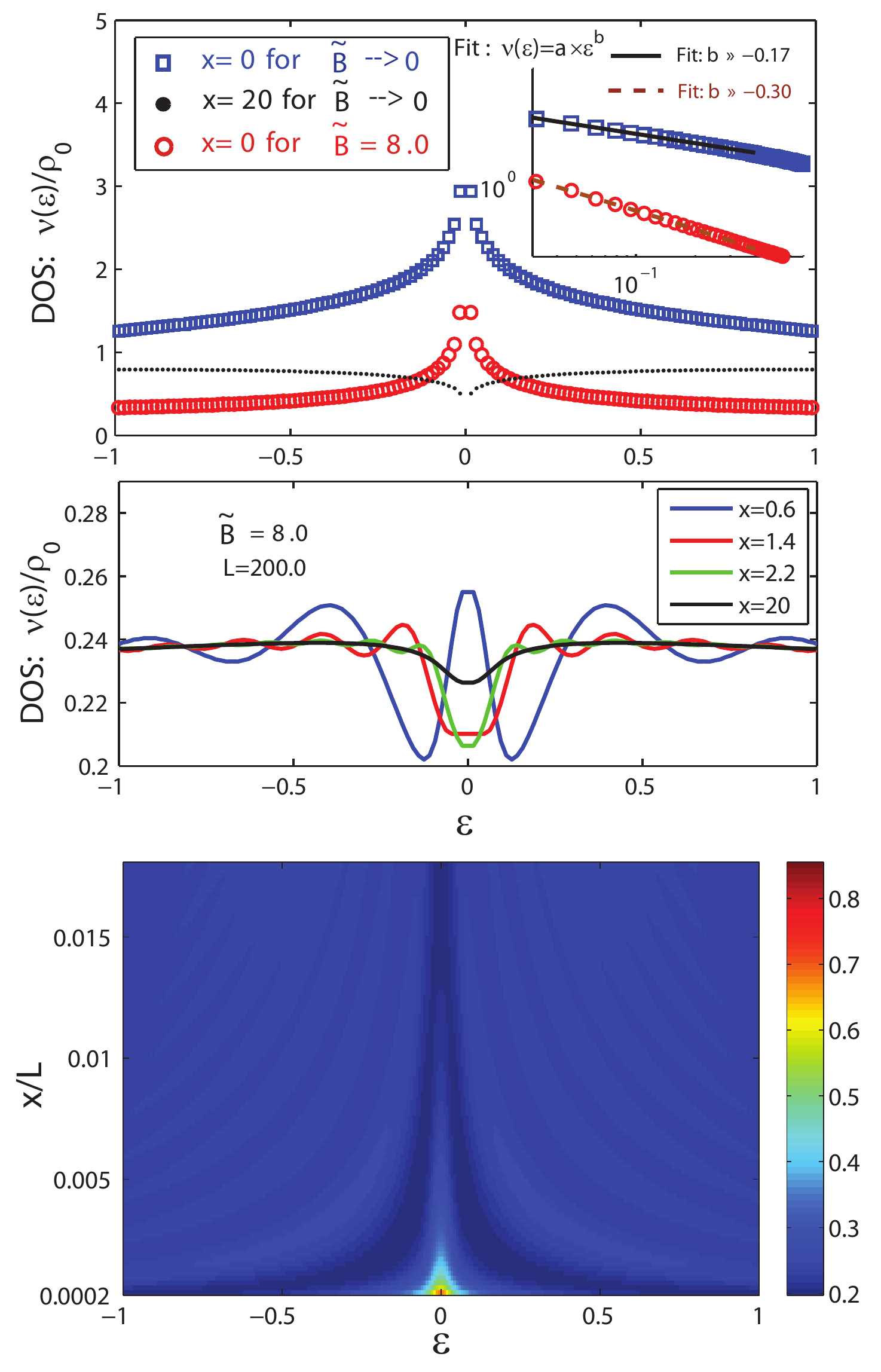}
\caption{Upper panel: The blue square curve corresponds to
$\widetilde{B}=\pi B/2\alpha\rightarrow 0$ and $x=0$ (we choose
$\widetilde{B}=10^{-7}$ in numerics), with $\nu(\varepsilon)\propto
\varepsilon^{-0.17}$, where the power exponent for the spinful LL is
$g_{\rho}/2+1/(2 g_{\sigma})-1=-0.162$. The black dotted curve is
for $\widetilde{B}\rightarrow 0$ and $x=20.0$. The red circle curve
is for $\widetilde{B}=8.0$ and $x=0$, with $\nu(\varepsilon)\propto
\varepsilon^{-0.3}$. The inset shows the fitting to the power law
expressions. Middle panel: TDOS at $x\neq 0$ from the SC-SLL
interface. Lower panel: Two-dimensional map of the TDOS in the
$x-\varepsilon$ plane. Parameters: $L=200.0$, $\alpha=0.04$,
$g_{\rho}=0.5$, $g_{\sigma}=0.85$, $v_{\rho}=0.4$,
$v_{\sigma}=0.235$, and $\chi=0$. The unit of TDOS is $L/2\pi$. }
\label{fig:DOS_SCHA}
\end{figure}

\section{Self-consistent harmonic approximation (SCHA)}

The TDOS can be also obtained by using SCHA, i.e., expanding the
cosine term in Eq.~\eqref{eq:H_wire-SLL} up to the second order
around one of the minima. Then, the effective Hamiltonian becomes
quadratic
\begin{eqnarray}
H_{\rm W} = \sum_{\nu=\rho,\sigma}  \frac{v_{\nu}}{2} \int_{0}^{L}
dx\big[g_{\nu} (\partial_x\theta_{\nu})^2
            +g^{-1}_{\nu}(\partial_{x}\phi_{\nu})^2\big]  \nonumber\\
 \quad +\frac{\pi B}{2\alpha} \int_{0}^{L} dx \big( \phi_{\rho}+\theta_{\sigma}-(2l+1)
 \sqrt{\pi/2}\big)^2.
\label{eq:H_SOC_eff_quad}
\end{eqnarray}
At this stage one inserts the mode expansion from
Eq.~(\ref{eq:ModeExpansion}) into Eq.~(\ref{eq:H_SOC_eff_quad}), and
diagonalizes the new Hamiltonian using the Bogolubov-Hopfield
transformation numerically (see Appendix~\ref{app:SCHA}). After the
diagonalization, one can obtain the time evolution of the mode
expansion in Eq. (\ref{eq:ModeExpansion}) in terms of their new
eigenmodes and eigenenergies. The TDOS defined in
Eqs.~(\ref{eq:DOSF}) and (\ref{eq:FC}) can then be computed
numerically.

In Fig.~\ref{fig:DOS_SCHA} we plot the resulting TDOS for a finite
wire ($L=200$) at the Rashba wire-superconductor interface $x=0$
(upper panel) and at $x=0.6, 1.4, 2.2, 20.0$ (lower panel) for
$g_{\rho}=0.5$ and $g_{\sigma}=0.85$ as a function of the energy
$\varepsilon$ (or equivalently bias voltage $eV$). The $B\rightarrow
0$ curve corresponds to the ordinary spinful Luttinger liquid, which
shows zero-bias enhancement $\nu(\varepsilon)\propto
\varepsilon^{g_{\rho}/2+1/(2g_{\sigma})-1}$ at the interfaces, which
is consistent with Ref.~\onlinecite{Winkelholz96}. In contrast,
farther away from the interface ($x=0.1 L$), the TDOS displays the
usual power law suppression at zero voltage bias. The curves for the
nonvanishing field are shown for the case that $\mu$ lies in the the
middle of the Zeeman gap $k_{F}=k_{\rm SO}$, i.e., SLL limit. At the
interface the TDOS for the SLL also exhibits an anomalous
enhancement at zero bias, but with a different power exponent [see
Eq.~(\ref{eq:ADOS_interface})]. The middle panel shows the TDOS at
$x=0.6, 1.4, 2.2, 20.0$ away from the SC-SLL boundary. The
zero-energy peak survives for small $x$ ($x=0.6$) and vanishes as
$x$ increases. The TDOS for $x=0.6, 1.4, 2.2$ shows the oscillations
and their amplitudes are reduced when increasing $x$. Those
signatures are consistent with the factor $x^{-\eta/2-3\beta/2}$ in
our analytical asymptotic result Eq.~(\ref{eq:DOS_FiniteX}). Because
of the suppression factor and the finite frequency resolution in
numerics (note that there is always such a frequency cutoff in
experiments), the oscillation disappears for large $x$ (e.g.,
$x=20.0$). A two-dimensional color map of the DOS near the SC-SLL
interface $x=0$ as both functions of $\varepsilon$ and $x$ is
plotted in the lower panel of Fig.~\ref{fig:DOS_SCHA}, which shows
the zero-bias enhancement near the SC-SLL interface, the zero-bias
suppression far away from $x=0$, and the Friedel-type oscillation.

\section{Summary}

We have studied tunneling density of states into a quantum wire with
strong spin-orbital coupling proximitized to superconductors. The
delicate interplay of superconducting correlations and Luttinger
liquid interactions leads to a dramatic change in the zero-bias
anomaly which transforms into a peak. This signature is a
consequence of the Andreev reflections at the SC-SLL interface. Our
predictions may trigger new experiments and can be tested in carbon
nanotubes~\cite{Kasumov,Morpurgo} or InAs quantum wires.~\cite{Vlad}
Perhaps it is plausible to argue that yet unexplained narrow
needlelike resonance pinned at zero bias of a superconductor-InAs
nanowire-superconductor device~\cite{Vlad} is in fact related to the
anomalous enhancement of the density of states in a wire due to
proximity effect and can be qualitative explained by our theory.

There is one important comment in order of the system under
consideration. If the wire is built on top of a superconductor, the
spiral Luttinger liquid, in the part that is proximitized to the
superconducting bulk, is in its topological superconducting phase
and thus supports Majorana fermions at the
interfaces~\cite{Gangadharaiah11,Sela11,Stoudenmire11}. In this
case, the zero-bias anomaly peak feature due to Andreev reflections,
discussed in this paper, coexists with the zero-bias peak due to the
Majorana fermion~\cite{law09,kouwenhovenSCI12,deng12,das12}. As the
chemical potential is tuned far above the Zeeman gap the zero-bias
anomaly peak due to Andreev reflection disappears, which also
coincides with the disappearance of Majorana fermions. Therefore,
our signature in the tunneling density of states masks the possible
presence of the Majorana fermion. This brings yet another important
detail that should be carefully looked at when interpreting
experimental data.

\subsection*{Acknowledgments}

D.E.L. was supported by Michigan State University and in part by ARO
through Contract No. W911NF-12-1-0235. A.L. acknowledges support
from NSF under Grant No. PHYS-1066293, and the hospitality of the
Aspen Center for Physics where part of this work was performed.

\appendix

\section{Derivation of TDOS in Eq.~(\ref{eq:ADOS_interface}) and (\ref{eq:DOS_FiniteX})}
\label{app:DOS}
The TDOS is given in terms of the time correlation function
\begin{equation}
 \nu(x,\varepsilon)=\!\!\int_{-\infty}^{+\infty}\!\! \frac{dt}{2\pi} \, e^{i\varepsilon t}
   \Big[ \langle \Psi(x,t) \Psi^{\dagger}(x,0) \rangle +\langle \Psi^{\dagger}(x,0)\Psi(x,t) \rangle  \Big],
\end{equation}
where $\langle \Psi^{\dagger}(x,0)\Psi(x,t) \rangle$ can be computed
similar to $\langle \Psi(x,t) \Psi^{\dagger}(x,0) \rangle$, and its
result is obtained by changing $t$ to $-t$ in
Eq.~(\ref{eq:corr_Linf}). At the SC-SLL interface $x=0$, the
correlation function yields
\begin{equation}
 \langle \Psi(0,t) \Psi^{\dagger}(0,0) \rangle= \frac{1}{\pi\alpha}\left[\frac{\alpha}{i u_{-}t + \alpha}
 \right]^{2\eta}.
\end{equation}
Fourier transforming this one finds Eq.~(\ref{eq:ADOS_interface}) of
the main text.

\begin{figure}
\centering
\includegraphics[width=3.0in,clip]{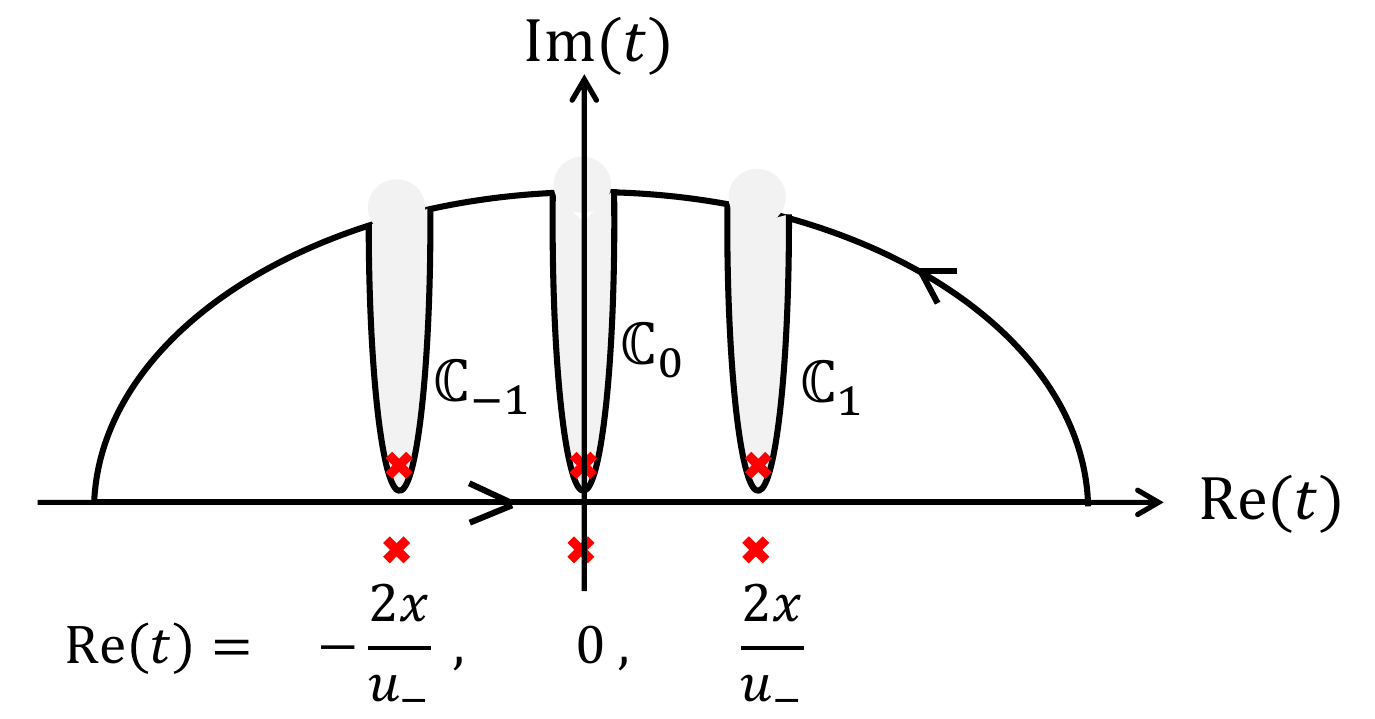}
\caption{Integration contour in the upper complex plane of $t$. The
red cross symbols are branching points for the correlation function
$\langle \Psi(x,t) \Psi^{\dagger}(x,0) \rangle$ (upper plane) and
$\langle \Psi^{\dagger}(x,0)\Psi(x,t) \rangle$ (lower plane). }
\label{fig:CountorIntegral}
\end{figure}

For $x\neq 0$, the TDOS: $\nu(x,\varepsilon>0)$ can be obtained by
integrating over $t$ along three branch cuts, i.e.
$\mathbb{C}_{-1}$, $\mathbb{C}_{0}$, and $\mathbb{C}_{1}$ (with
branching points $i\alpha$ and $\pm2x/u_{-}+i\alpha$), with
integrand $\langle \Psi(x,t) \Psi^{\dagger}(x,0) \rangle$ in the
upper complex $t$ plane as shown in Fig.~\ref{fig:CountorIntegral}.
The $\nu(x,\varepsilon<0)$ is obtained by integrating over $t$ along
three other branch cuts with integrand $\langle
\Psi^{\dagger}(x,0)\Psi(x,t) \rangle$ in the lower complex $t$
plane. Let us focus on the $\varepsilon>0$ case,
\begin{eqnarray}
 &&\nu(x,\varepsilon>0)=\frac{1}{2\pi^2\alpha} \int_{\mathbb{C}_{-1} +\mathbb{C}_{0}+\mathbb{C}_{1}} dt
   e^{i\varepsilon t} \left[\frac{\alpha}{i u_{-}t + \alpha}\right]^{\eta +
   \beta}\nonumber\\
  &&\times\left[\frac{\alpha^2+(2x)^2}{[i(u_{-}t-2x)+\alpha][i(u_{-}t+2x)+\alpha]}\right]^{\frac{\eta-\beta}{2}}\nonumber\\
  &&=I_{\mathbb{C}_{-1}}+ I_{\mathbb{C}_{0}}+ I_{\mathbb{C}_{1}}.
\end{eqnarray}
In the limit $2x\varepsilon/u_{-}\gg 1$, the contribution of those
branch cuts can be calculated independently. First of all, the
integral $I_{\mathbb{C}_{0}}$ is
\begin{eqnarray}
 I_{\mathbb{C}_{0}}\approx \frac{1}{2\pi^2\alpha} \left[ \frac{\alpha^2+(2x)^2}{(2x)^2} \right]^{\frac{\eta-\beta}{2}}
      \int_{\mathbb{C}_{0}} dt e^{i\varepsilon t} \left[\frac{\alpha}{i u_{-} t}\right]^{\eta +
      \beta}\nonumber\\
  =\frac{1}{\pi \Gamma(\eta+\beta)u_{-}}\left[ \frac{\alpha \varepsilon}{u_{-}}\right]^{\eta+\beta-1}
\end{eqnarray}
as $x\gg\alpha$. Second, the integral involving the path
$I_{\mathbb{C}_{-1}}$ can be simplified by using a variable
substitution $\tau=t+2x/u_{-}$,
\begin{eqnarray}
\hskip-.45cm &&I_{\mathbb{C}_{-1}}=\frac{1}{2\pi^2\alpha}
\alpha^{\eta+\beta} (\alpha^2+4x^2)^{\frac{\eta-\beta}{2}}
     e^{-i\frac{2x\varepsilon}{u_{-}}}\nonumber\\
\hskip-.45cm &&\times\!\!\int_{\mathbb{C}_{0}}\!\!\! d\tau
e^{i\epsilon\tau}\!\!
     \left[\frac{1}{i (u_{-}\tau-2x)} \right]^{\eta+\beta}\!\!
     \left[\frac{1}{i
     (u_{-}\tau-4x)}\right]^{\frac{\eta-\beta}{2}}\!\!
     \left[\frac{1}{i u_{-}\tau}\right]^{\frac{\eta-\beta}{2}} \nonumber\\
\hskip-.45cm
&&=\frac{\alpha^{\eta+\beta-1}i^{\frac{\eta-\beta}{2}}}{\pi\Gamma(\frac{\eta-\beta}{2})u_{-}}
\left[\frac{i}{2}\right]^{\eta+\beta}\!\!
    \left[\frac{\varepsilon}{u_{-}}\right]^{\frac{\eta-\beta}{2}-1}\!\!\!
      x^{-\frac{\eta}{2}-\frac{3\beta}{2}}
      e^{-i\frac{2x\varepsilon}{u_{-}}}.
\end{eqnarray}
Similarly, the last integral involving the path $I_{\mathbb{C}_{1}}$
can be simplified by using a variable substitution
$\tau=t-2x/u_{-}$, and then
\begin{equation}
 I_{\mathbb{C}_{1}}=\frac{\alpha^{\eta+\beta-1}(-i)^{\frac{\eta-\beta}{2}}}{\pi\Gamma(\frac{\eta-\beta}{2})u_{-}}
 \left[\frac{-i}{2}\right]^{\eta+\beta}\left[\frac{\varepsilon}{u_{-}}
\right]^{\frac{\eta-\beta}{2}-1}\!\!\!
      x^{-\frac{\eta}{2}-\frac{3\beta}{2}}
      e^{i\frac{2x\varepsilon}{u_{-}}}.
\end{equation}
Summing up all the terms, one can obtain the TDOS asymptote, i.e.,
Eq.~(\ref{eq:DOS_FiniteX}) from the main text.

\section{Diagonalization of effective Hamiltonian in SCHA}
\label{app:SCHA}

Inserting the mode expansion from
Eq.~(\ref{eq:ModeExpansion}) into Eq.~(\ref{eq:H_SOC_eff_quad}), we get
\begin{eqnarray}
H &=& \frac{g_{\rho} v_{\rho}(J+\chi)^2}{4L} +\frac{v_{\sigma}
M^2}{4L g_{\sigma}}
   + \frac{\widetilde{B}L}{\pi} ( \hat{\phi}_{\rho}^{0}+\hat{\theta}_{\sigma}^{0}-U_{min})^2\nonumber\\
&& +\sum_{q>0}\Big(  \sum_{\nu=\rho,\sigma}\frac{v_{\nu}}{2}q
     (2 b_{\nu q}^{\dagger}b_{\nu q}+1) \nonumber\\
&&     + \frac{\widetilde{B} g_{\rho}}{2 q}(b_{\rho q}^{\dagger}+b_{\rho q})^2
+\frac{\widetilde{B}}{2 g_{\sigma} q}(b_{\sigma q}^{\dagger}+b_{\sigma q})^2  \nonumber\\
 &&  +\frac{\widetilde{B}}{q}\sqrt{\frac{g_{\rho}}{g_{\sigma}}}
   (b_{\rho q}^{\dagger}+b_{\rho q})(b_{\sigma q}^{\dagger}+b_{\sigma q})\Big),
\label{eq:H_eff_ME}
\end{eqnarray}
where $U_{min}=(2l+1)\sqrt{\pi/2}$.
We can apply a canonical transformation to the topological part:
\begin{eqnarray}
 \Phi_{1} &=& \hat{\phi}_{\rho}^{0}+\hat{\theta}_{\sigma}^{0}-U_{min} \;,\nonumber\\
N_{1}&=&[(J+\chi)+\kappa M]/(1+\kappa) \;,\nonumber\\
\Phi_{2} &=& \sqrt{\kappa}\hat{\phi}_{\rho}^{0}-\hat{\theta}_{\sigma}^{0}/\sqrt{\kappa}\;,\nonumber\\
 N_{2}&=&\sqrt{\kappa}[(J+\chi)- M]/(1+\kappa)\;,
\end{eqnarray}
with $\kappa=v_{\sigma}/(v_{\rho}g_{\rho}g_{\sigma})$. By further
introducing ladder operators $\eta_{1}$ and $\eta_{1}^{\dagger}$,
\begin{eqnarray}
 \Phi_{1}=\Xi^{-1/4}(\eta_{1}+\eta_{1}^{\dagger})/\sqrt{2} \;,\nonumber\\
  N_{1}= i \;\Xi^{1/4}(\eta_{1}^{\dagger}-\eta_{1})/\sqrt{2},
\end{eqnarray}
with
$\Xi=4\widetilde{B}L^2/(\pi(g_{\rho}v_{\rho}+v_{\sigma}/g_{\sigma}))$,
the topological part of Eq.~(\ref{eq:H_eff_ME}) is reduced to
\begin{equation}
 H_{\rm TOPO}=\sqrt{\frac{\widetilde{B}}{\pi}(g_{\rho}v_{\rho}+\frac{v_{\sigma}}{g_{\sigma}})}
                \Big( \eta_{1}^{\dagger}\eta_{1}-\frac{1}{2} \Big)
           +\frac{g_{\rho}v_{\rho}+\frac{v_{\sigma}}{g_{\sigma}}}{4L}N_{2}^2.
\end{equation}
The non topological excitations can be diagonalized using the
Bogolubov-Hopfield transformation, and we will briefly outline the
main procedures below. This term can be diagonalized:
\begin{eqnarray}
 H_{\rm NT}&=&\sum_{q>0} \vec{b}_{q}\cdot H_{\mathrm{NT},q}\cdot \vec{b}_{q}^{T}\nonumber\\
   &=&\sum_{q>0} \vec{c}_{q}\cdot \rm{Diag}\{E_{1q},E_{2q},E_{1q},E_{2q}\}\cdot \vec{c}_{q}^{T}\;,
\vspace{0.02in}
\end{eqnarray}
where $\vec{b}_{q}=(b_{\rho q}^{\dagger},b_{\sigma
q}^{\dagger},b_{\rho q},b_{\sigma q})$ and the eigenvector after
diagonalization is $\vec{c}_{q}=(c_{1 q}^{\dagger},c_{2
q}^{\dagger},c_{1 q},c_{2 q})$. The transformation matrix
$\textbf{Q}$, i.e., $\vec{b}^{T}=\textbf{Q}\cdot \vec{c}^{T}$, can
be obtained by the relation $\textbf{Q}=K \cdot M^{\dagger}\cdot K$,
where $K=\rm{Diag}\{\mathbb{I}_{2\times 2},-\mathbb{I}_{2\times 2}
\}$. The matrix $M^{\dagger}$ is obtained by solving the eigenvalue
problem:
\begin{equation}
 (H_{\mathrm{NT},q}K) M^{\dagger}=M^{\dagger} \rm{Diag}\{E_{1q},E_{2q},-E_{1q},-E_{2q}\}.
\end{equation}
One can simply diagonalize the non topological part numerically.
After the diagonalization, one can obtain the time evolution of the
mode expansion for Eq.~(5) of the main text in terms of their new
eigenmode and eigenenergies. The TDOS is then computed numerically.

\end{document}